\documentclass[aps,prl,twocolumn,superscriptaddress,longbibliography]{revtex4-1}
\usepackage{blindtext}
\usepackage{amssymb}
\usepackage{amsmath}
\usepackage{graphicx}
\usepackage[caption=false]{subfig}
\usepackage{floatrow}
\usepackage[dvipsnames]{xcolor}
\usepackage[version=3]{mhchem}
\usepackage{footnote}
\usepackage{bm}
\usepackage{dsfont}
\usepackage{xr-hyper}
\usepackage[hidelinks]{hyperref}
\usepackage{cleveref}
\usepackage{soul}
\usepackage[normalem]{ulem}

\usepackage{hhline}

\usepackage{booktabs}
\crefname{equation}{Eq.}{Eqs.}
\Crefname{equation}{Equation}{Equations}
\crefname{figure}{Fig.}{Figs.} 
\Crefname{figure}{Figure}{Figures}
\crefname{section}{Sect.}{Sects.}
\Crefname{section}{Section}{Sections}
\crefname{table}{Table}{Tables}
\crefname{appsec}{}{Appendices}

\newcommand{\ket}[1]{\ensuremath{\left|#1\right\rangle}}
\newcommand{\bra}[1]{\ensuremath{\left\langle#1\right|}}

\newcommand{\ketbra}[2]{\ensuremath{\left|#1\right\rangle\left\langle#2\right|}}

\newcommand{\matrixel}[3]{\ensuremath{\left\langle#1|#2|#3\right\rangle}}

\definecolor{catcolor}{rgb}{0.35, 0.05, 0.75}
\definecolor{catcolorcritical}{rgb}{0.1, 0.5, 0.3}
\definecolor{mariecolor}{rgb}{0.65,0.4,0.1}

\makeatletter
\def\blfootnote{\xdef\@thefnmark{}\@footnotetext}
\makeatother

\usepackage{tabularx}

\begin{document}

\floatsetup[figure]{style=plain,subcapbesideposition=top}

\title{Cat-qubit-inspired gate on cos($2\theta$) qubits}

\author{Catherine Leroux}
\email{Catherine.Leroux@USherbrooke.ca}
\thanks{This work was done prior to joining AWS.}
\affiliation{Institut quantique \& D\'epartement de Physique, Universit\'e de Sherbrooke, Sherbrooke J1K 2R1 QC, Canada}

\author{Alexandre Blais}
\affiliation{Institut quantique \& D\'epartement de Physique, Universit\'e de Sherbrooke, Sherbrooke J1K 2R1 QC, Canada}
\affiliation{Canadian Institute for Advanced Research, Toronto, ON, Canada}

\date{\today}

\begin{abstract}
For $\cos(2\theta)$ qubits based on voltage-controlled semiconductor nanowire Josephson junctions we introduce  a single-qubit $Z$ gate inspired by the noise-bias preserving gate of the Kerr-cat qubit. This scheme relies on a $\pi$ rotation in phase space via a beamsplitter-like transformation between a qubit and ancilla qubit. The rotation is implemented by adiabatically changing the potential energies of the two qubits such as to preserve a double-well potential at all times. This gate constrains the dynamics in the subspace of a $\cos(2\theta)$ qubit at all times, therefore yielding high-fidelity operation while preserving the qubit's coherence.
We introduce a circuit to realize this gate and support our findings with numerical simulations.
\end{abstract}
                 
\maketitle

\emph{Introduction---} The protection mechanism of the $\cos(2\theta)$ qubit relies on the disjoint support of the qubit's logical eigenstates~\cite{Gyenis2021_moving,Smith2020}. This protection mechanism, however, also makes transitions between the logical states necessary for gates difficult. Realizing gates on protected qubits therefore rely on breaking the symmetry of the qubit's potential wells or on the use of higher energy states. 
Both approaches expose the qubit to additional decay mechanisms, and often result in slow gates~\cite{Gyenis2021,Paolo_2019}. 
Ideally, high-fidelity gates would maintain the qubit's double-well structure, thereby preserving its underlying protection, while also being fast.

The double-well structure of the potential energy of the 
$\cos(2\theta)$ qubit is reminiscent of the metapotential of the Kerr-cat qubit which encodes logical states in coherent states of opposite phases. In this qubit, those states are stabilized by opposing the Kerr nonlinearity of a Josephson junction with a two-photon pump~\cite{Puri2017,Grimm2020}. An advantage of this qubit is its exponentially long bit-flip time with the amplitude of the coherent states.
Importantly, the two-photon drive in the Kerr-cat qubit not only stabilizes the qubit but is also a resource when realizing gates. Indeed, by adiabatically changing the phase of the drive, it is possible to implement a fast, high-fidelity $\pi$ rotation of the cat states in phase space such that the two metapotential wells of the Kerr cat are swapped leading to the desired operation~\cite{Puri2020}. Because the logical states remain confined by the metapotential throughout the gate, bit-flip errors remain exponentially suppressed during the protocol.

Inspired by this approach, here we show how to realize a logical $Z$ gate on a $\cos(2\theta)$ qubit based on voltage-controlled semiconductor nanowire Josephson junctions~\cite{Larsen2020,Schrade2022}. 
This qubit is formed by a capacitively shunted flux-biased interferometer made from voltage-controlled semiconductor nanowire Josephson junctions. Crucially, as experimentally demonstrated in Ref.~\cite{Larsen2020}, this design allows to adiabatically shape the potential energy of the qubit between a single-well and a two-well potential by using the voltage bias. 

The central idea of the gate introduced here is to exploit this feature to perform an adiabatic manipulation of the energy potential of two semiconducting qubits, where one is the logical unit and the other is an ancillary qubit, such as to effectively rotate the $\cos(2\theta)$ potential in phase space resulting in a $Z$ gate on the logical qubit. Although we focus on the semiconducting $\cos(2\theta)$ qubit, these ideas can be extended to other 
protected qubits having a double-well potential structure by supplementing that qubit with a second mode. As will be made clear below, this is possible as long as both modes can have their potential energies varied between $\cos(\theta)$ and $\cos(2\theta)$. Such qubits include the fluxonium~\cite{Manucharyan2009} and other compact $\cos(2\theta)$ qubits~\cite{Smith2020,Paolo_2019}. 

This article is organized as follows: We review the main properties of the superconductor-semiconductor $\cos(2\theta)$ qubit before introducing the concept of the gate. We then discuss a possible circuit implementation and present numerical results before concluding.

\begin{figure}[t!]
    \centering   \includegraphics[width=\textwidth]{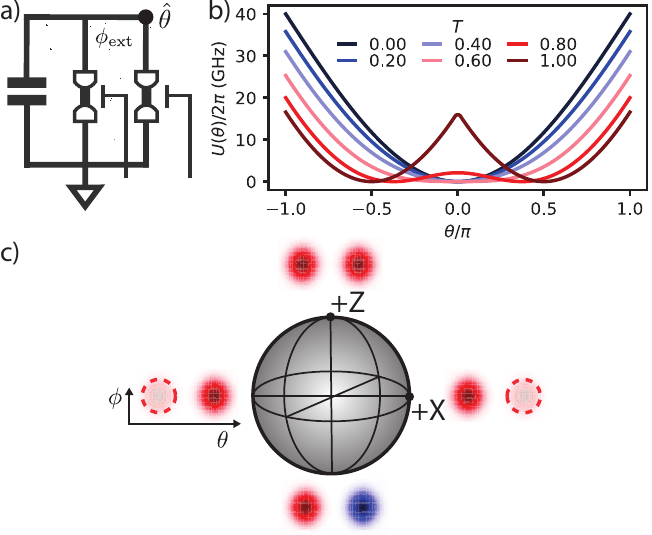}
    \caption{a) Circuit of a superconductor-semiconductor $\cos(2\theta)$ qubit composed of two capacitively shunted semiconducting junctions in parallel, biased at half-flux quantum. b) Potential energy of the $\cos(2\theta)$ qubit assuming a single transmission channel in each of the junctions with an identical superconducting gap $\Delta/h = 40$ GHz. The transmission probability is fixed to $T=1$ in the left junction and $T$ is varied in the right junction. c) Bloch sphere of a logical qubit with its logical states defined and illustrated in the 2D phase space of the two qubits with phase operators $\theta$ and $\phi$. Because these two operators commute, there are no interference fringes.}
    \label{fig:cos2theta}
\end{figure}

\emph{Superconductor-semiconductor $\cos(2\theta)$ qubit---} A superconductor-semiconductor $\cos(2\theta)$ qubit is a capacitively shunted SQUID made of two semiconducting junctions which we take to be flux-biased at half-quantum flux, see \cref{fig:cos2theta} a)~\cite{Larsen2020,Schrade2022}. Each junction can be biased with a gate voltage to control the junction's effective Josephson energy. The qubit is described by the approximate Hamiltonian (see \cref{app:cos(2theta) qubit} for details on the derivation of this model)
\begin{equation}
\begin{split}
    \hat H_{\cos(2\theta)} & \approx 4 E_C\hat n_\theta^2  - \alpha \cos(\hat\theta) + \beta \cos(2\hat \theta)
    , 
\end{split}\label{eq: h 2theta}
\end{equation}
where $\hat \theta$ is its phase operator with canonical number operator $\hat n_\theta$, $E_C$ is the charging energy, $\alpha =E_{J1}^1-E_{J1}^2$ and $\beta =  E_{J2}^1 + E_{J2}^2$ where $E_{Jk}^i(V_i)$ is the amplitude of the $k$th harmonic of the Andreev bound state energy of the $i$th junction gated with a voltage $V_i$. For  conciseness, the voltage dependence is not made explicit in \cref{eq: h 2theta}. As noted in Ref.~\cite{Larsen2020}, the junction asymmetry can be used to constrain the qubit to a single potential well (for ${|E_{J1}^1-E_{J1}^2| > |E_{J2}^1+E_{J2}^2|}$) or to two potential wells (for ${|E_{J1}^1-E_{J1}^2| < |E_{J2}^1+E_{J2}^2|}$), see \cref{fig:cos2theta} b).

\emph{Concept---} Realizing a logical $Z$ gate in a $\cos(2\theta)$ qubit  without breaking the qubit's protection can be challenging. Our approach to achieve this is analogous to the noise bias preserving $Z$ gate of the Kerr-cat qubit~\cite{Puri2017} where the cat's metapotential is adiabatically rotated in phase space to perform the gate~\cite{Puri2020}. With the Kerr-cat qubit, this is realized by adiabatically changing the phase of the qubit's two-photon pump. 

Inspired by this, we use a similar approach where now the control knob activating the gate is a beamsplitter angle between two modes, one having a $\cos(2\theta)$ potential and an ancilla mode with a $\cos(\phi)$ potential. As will become apparent below, the beamsplitter operation effectively swaps the double-well and single-well potential of the two modes in time, resulting in a rotation of the 
potential wells in the 2D plane spanned by the phase coordinates of the two modes. As in the Kerr-cat qubit, this rotation realizes a $Z$ gate.

To better understand this dynamic, let us first consider an idealized situation starting with the Hamiltonian
\begin{equation}
\begin{split}
    \hat H_0 &= 4 E_{C\theta}\hat n_\theta^2 + \beta \cos(2\hat{\theta}) + 4 E_{C\phi}\hat n_\phi^2- \alpha \cos(\hat{\phi})
\end{split}, \label{eq: free Hamiltonian}
\end{equation}
describing two non-interacting modes, with the $\hat\theta$ mode being a $\cos(2\hat\theta)$ qubit and the $\hat\phi$ mode being a transmon qubit. Here, $\beta$ and $\alpha$ are the amplitudes of the respective potential energy, and $E_{Cx}$ is the charging energy of the mode ${x\in\{\theta,\phi\}}$. 

\begin{figure*}[t!]
    \centering
    \includegraphics[width=\textwidth]{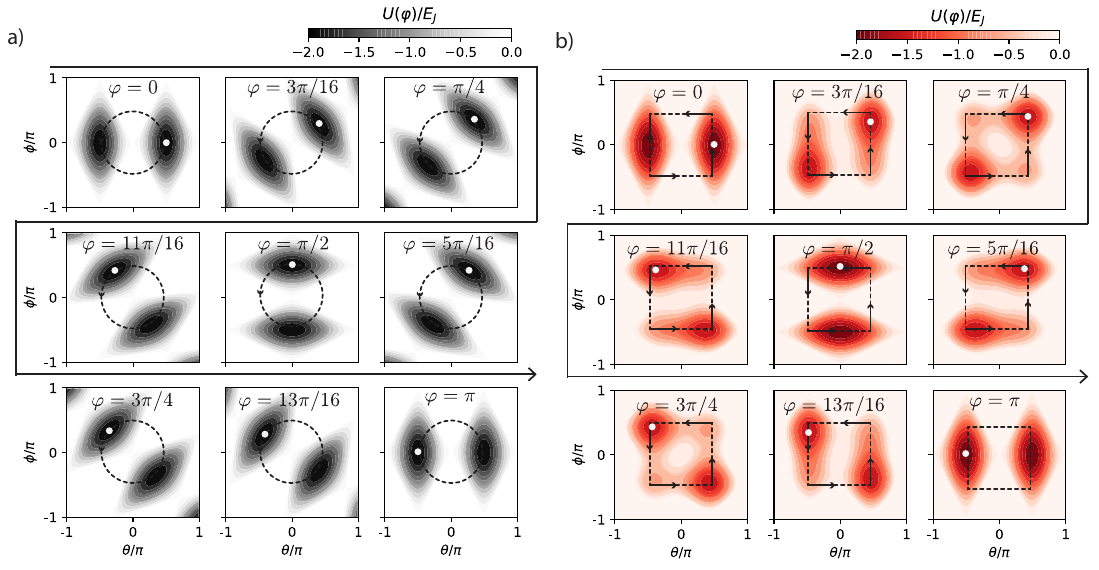}
    \caption{a) Rotation of the potential landscape in the 2D phase space defined by the two phase coordinates $\theta$ and $\phi$, obtained from the beamsplitter operation of \cref{eq: beamsplittered rotation}. The white dot tracks the rotation of one of the two potential wells. b) Rotation resulting from the protocol in \cref{eq: effective rotation}. This relies on the implementation of a large $\sin(\hat\theta)\sin(\hat\phi)$. Here $\beta=\alpha=\zeta=E_J$ for simplicity and the angles are, following the black arrow, $\varphi = 0$, $3\pi/16$, $\pi/4$, $5\pi/16$, $\pi/2$, $11\pi/16$, $3\pi/4$, $13\pi/16$ and $\pi$.  
    In b) the wells follow a square-like path instead of a circular path as in a). The $\theta$ mode, initially in a double-well potential, slowly goes through a single well potential and then back to a double-well potential while the $\phi$ mode does the opposite.}
    \label{fig: zak patch rotation vs ideal}
\end{figure*}

The logical states of this two-mode system are also illustrated in \cref{fig:cos2theta}~c) in the 2D phase space of the two commuting observables $\{\hat\theta,\hat\phi\}$ with the convention that the even and odd parity states are along the $Z$ axis. Similarly to cat qubits, the logical states are exponentially protected against $Z$-type errors rendering any $Z$ gate challenging. The $Z$ gate can, however, be robustly realized via a rotation in the 2D phase space while preserving this protection, again in analogy to the rotation in the phase space of cat qubits.

In principle, a rotation of the $\theta$ and $\phi$ coordinates can be implemented with the beamsplitter transformation $\hat B(\varphi)$ defined such that 
\begin{align}
    &\hat B\hat{\theta}\hat B^\dagger = \cos(\varphi) \hat{\theta} - \sin(\varphi) \hat{\phi}, \\ 
    &\hat B\hat{\phi}\hat B^\dagger = \cos(\varphi) \hat{\phi}+ \sin(\varphi) \hat{\theta}, 
\end{align}
where $\varphi$ is the beamsplitter angle and therefore the angle of rotation of the $\theta$ and $\phi$ coordinates. Under this transformation the potential energy ${\hat U = \beta\cos(2\hat{\theta})-\alpha\cos(\hat{\phi})}$ of the Hamiltonian $\hat H_0$ transforms as 
\begin{equation}
\begin{split}
    \hat B\hat U \hat B^\dagger = 
    \beta\cos\left(2[\cos(\varphi) \hat{\theta} +\sin(\varphi)\hat{\phi}]\right) \\
    - \alpha\cos\left([\cos(\varphi) \hat{\phi}-\sin(\varphi) \hat{\theta}]\right). \label{eq: beamsplittered rotation}
\end{split}
\end{equation}
The rotation of the double-well potential under this transformation is illustrated in \cref{fig: zak patch rotation vs ideal} a) for angles $\varphi \in [0,\,\pi]$. As desired, the two potential well along $\theta$ at $\varphi = 0$  exchange their place after a $\varphi = \pi$ rotation. This simple observation suggests that a $Z$ gate can be realized by adiabatically rotating the potential energy, if implementing \cref{eq: beamsplittered rotation} was possible. 

Since a transformation of the form of \cref{eq: beamsplittered rotation} appears to be difficult in practice, as an alternative we propose to approximate this operation by relying on a tunable potential energy of the form
\begin{equation}
\begin{split}
    \hat U_\varphi & = 
    \beta \cos^2(\varphi)\cos(2\hat{\theta}) - \alpha\cos^2(\varphi) \cos(\hat{\phi}) \\
    & + \beta\sin^2(\varphi) \cos(2\hat{\phi}) - \alpha \sin^2(\varphi) \sin(\hat{\theta}) \\
    & - \frac{\zeta}{2} \sin(2\varphi) \sin(\hat{\theta})\sin(\hat{\phi}),
\end{split} \label{eq: effective rotation}
\end{equation}
which we will show can be engineered.  In this expression, $\zeta$ is the interaction strength between the two qubits, to be optimized and whose sign controls the direction of the rotation. The term proportional to $\zeta$ in \cref{eq: effective rotation} breaks the fourfold degeneracy of the potential of \cref{eq: free Hamiltonian} such as to preserve a double-well structure at all times. 
 
The corresponding rotation of the $\theta$ and $\phi$ coordinates is illustrated in \cref{fig: zak patch rotation vs ideal} b) for different angles $\varphi$. We observe that \cref{eq: effective rotation} yields the desired rotation similar to \cref{eq: beamsplittered rotation} but on a square instead of a circle in the 2D phase space.

Just as in the previous example, the potential wells along $\theta$ are exchanged under this operation suggesting that a $Z$ gate can be realized by adiabatically swapping the roles of the $\theta$ and $\phi$ modes, i.e.~by slowly going back and forth between a double-well potential and a single well potential in each mode. The interaction $\zeta$ is used here to control the direction of the rotation. In contrast to the Kerr-cat qubit~\cite{Puri2017} where the wells move along a circle, here the wells approximately move along the edges of a square, see \cref{fig: zak patch rotation vs ideal} b). As a consequence, the distance $d$ between the two global minimum increases during the rotation, i.e.~${d = \pi |\sec(\min (\varphi,\pi/2-\varphi)|}$ varies between $\pi$ and $\sqrt{2}\pi$ (with $0\leq \varphi \leq \pi/2$). This small change in the distance between the potential wells has little effect on the qubit's protection.

Below, we show how realize this concept by exploiting the fact that the superconductor-semiconductor qubit of \cref{fig:cos2theta} a) can be continuously tuned from having a $\cos(2\theta)$ potential to a $\cos(\theta)$ qubit by varying the transmission probability in one of its two junctions~\cite{Larsen2020}.

\begin{figure}[t!]
    \centering
    \includegraphics[width=.9\textwidth]{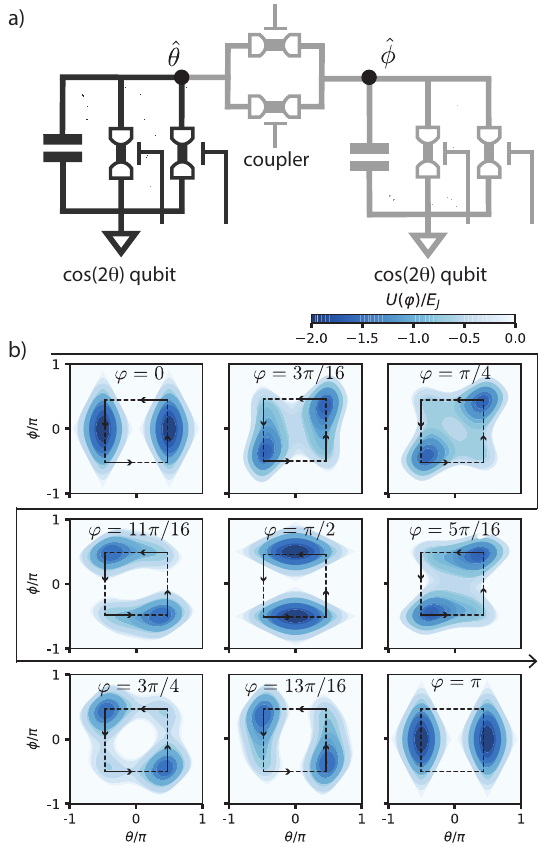}
    \caption{a) Possible circuit realization for a noise-preserving logical $Z$ gate in a superconductor-semiconductor $\cos(2\theta)$ qubit (in black). The coupler is used to implement a voltage-tunable $\cos(\hat{\theta}-\hat{\phi})$ interaction with the goal of approximately replicating \cref{fig: zak patch rotation vs ideal} b) at low energies. b) Potential energy of \cref{eq: circuit hamiltonian}, the Hamiltonian of the circuit in panel a), at $\varphi = 0$, $3\pi/16$, $\pi/4$, $5\pi/16$, $\pi/2$, $11\pi/16$, $3\pi/4$, $13\pi/16$ and $\pi$ in \cref{eq: effective rotation}.}
    \label{fig:Zak patch rotation}
\end{figure}

\emph{Circuit implementation---} The adiabatic change between a double-well potential and a single-well potential corresponding to the first two lines of \cref{eq: effective rotation} can be readily realized by tuning the gate voltages at the junctions of the superconductor-semiconductor $\cos(2\theta)$ qubits such as to vary the asymmetry between the energies of the junctions. The main challenge is implementing the ideally large $\sin(\hat{\theta})\sin(\hat{\phi})$ interaction that varies in time  corresponding to the last line of \cref{eq: effective rotation}. Here, we propose to use another superconductor-semiconductor interferometer, with small transmission junctions and flux-biased at half quantum flux, as a coupler implementing an interaction of the type $\cos(\hat \theta- \hat \phi)$ with an amplitude $\zeta$ controlled by the asymmetry between the interferometer's two junctions, see \cref{fig:Zak patch rotation} a) and \cref{app: circuit implementation} for details of the derivation. Expanding the above cosine term, this approach yields both the desired $\sin(\hat \theta)\sin(\hat \phi)$ contribution as well as an additional error term $\cos(\hat \theta)\cos(\hat \phi)$. The impact of this error term is analyzed further below.

The circuit of \cref{fig:Zak patch rotation} a), where the circuits of the two $\cos(2\theta)$ qubits are assumed here to be identical  
for simplicity, is described by the Hamiltonian 
\begin{equation}
\begin{split}
    \hat H &= 4 E_C\hat n_\theta^2+ \beta \cos^2(\varphi) \cos(2\hat{\theta})- \alpha \sin^2(\varphi) \cos(\hat{\theta})   
   \\
   &+ 4 E_C\hat n_\phi^2- \alpha\cos^2(\varphi)\cos(\hat{\phi})+ \beta\sin^2(\varphi) \cos(2\hat{\phi}) \\
   &- \frac{\zeta}{2}\sin(2\varphi)\cos (\hat\theta-\hat\phi),
\end{split} \label{eq: circuit hamiltonian}
\end{equation}
where $\alpha \sin^2(\varphi)$ and $\alpha \cos^2(\varphi)$ are the amplitudes of the first harmonic of the potential energy at an angle $\varphi$ controlled by the voltage bias for modes $\theta$ and $\phi$, respectively, while $\beta \cos^2(\varphi)$ and $\beta \sin^2(\varphi)$ are the amplitudes of the second harmonic, and $\zeta$ is the interaction strength. We take both modes to have charging energy $E_C$. We however note that the charging energies and the amplitudes of the first and second harmonics of the Andreev bound state energy do not need to be strictly the same for both junctions. The potential energy landscape of \cref{eq: circuit hamiltonian} illustrated in \cref{fig:Zak patch rotation} b) approximately reproduces \cref{fig: zak patch rotation vs ideal} b) preserving its essential features. 
In particular, we note that the total parity of the two modes $\theta$ and $\phi$, $(\theta,\phi)\to (-\theta,-\phi)$, is a symmetry of the Hamiltonian in \cref{eq: circuit hamiltonian} for all angles $\varphi$. 

Importantly, we find that the instantaneous Hamiltonian corresponding to a value of $\varphi$ can be approximated at low energies by the effective model (see \cref{app: low energy})
\begin{equation}
    \hat H \approx 4 E_C\hat n_\Theta^2 + \beta \cos(2\pi \hat \Theta/d) \\
    + 4 E_C\hat n_\Phi^2 - \alpha \cos(d\hat \Phi/\pi),
    \label{eq: low energy H}
\end{equation}
where $\hat \Theta = \cos(\varphi)\hat \theta + \sin(\varphi)\hat \phi$ is defined along the radius of the rotation, $\hat \Phi = \cos(\varphi)\hat \phi - \sin(\varphi)\hat \theta$ is its perpendicular coordinate, and $d(\varphi) = \pi |\sec(\min (\varphi,|\pi/2-\varphi|,|\pi-\varphi|)|$ is the instantaneous distance between the two global minima. Here, the logical states are constrained to a double-well potential at all times during the $Z$ gate with $\hat \Theta$ being the logical mode. As before, we note that the main consequence of following a square path instead of a circular path in the 2D phase space is the renormalization of the distance between the global potential minima which results in a small renormalization of the qubit frequency as well as the matrix elements of the noise operators.

\begin{figure}[t!]
    \centering
    \includegraphics[width=.9\textwidth]{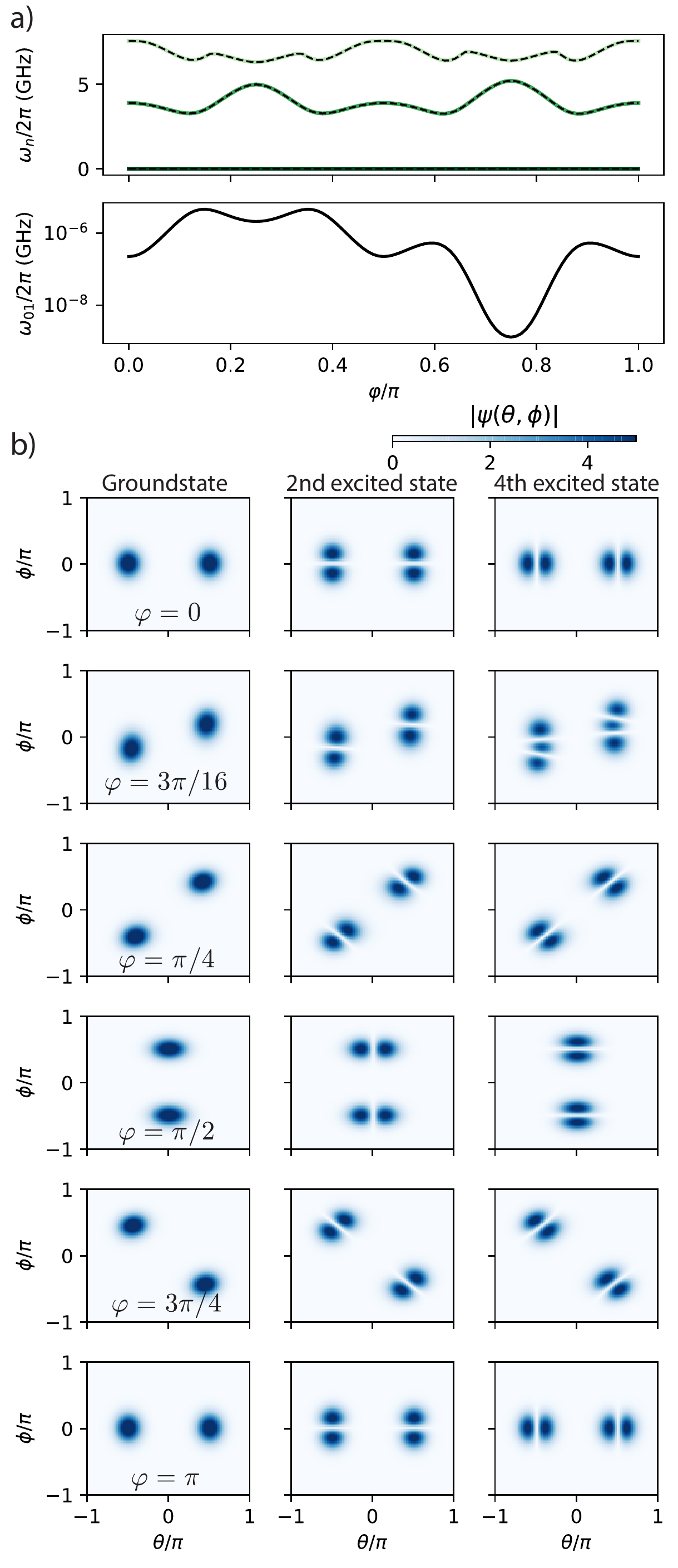}
    \caption{a) Energy spectrum of \cref{eq: circuit hamiltonian}, where $\omega_n$ is the $n$th eigenvalue, as a function of $\varphi/\pi$ for $E_C/h = 0.1$ GHz and $\alpha/h=\beta/h=\zeta/h=20.0$ GHz. Top panel: the $n$th green line corresponds to the eigenvalue $\omega_{2n}$ and the black line on top identifies the eigenvalue $\omega_{2n+1}$. Bottom panel: qubit 0-1 transition frequency in log scale. b) Eigenstates of \cref{eq: circuit hamiltonian} for different angles $\varphi$. Each column corresponds to an eigenstate 
    and each row is associated with a specific $\varphi$ as identified in the left column. Each panel shows the eigenstate in phase space as a function of $\theta/\pi$ and $\phi/\pi$. }
    \label{fig:spectrum}
\end{figure}

Going back to the full model of \cref{eq: circuit hamiltonian}, the first 6 energy levels versus $\varphi$ are illustrated in \cref{fig:spectrum} a) [top panel] as well as the 0-1 transition frequency [bottom panel] for $E_C/h = 100$ MHz and $\alpha/h=\beta/h=\zeta/h=20$~GHz. The eigenstates of even parity (for conciseness) are shown in panel b) for different values of $\varphi$. 
We observe that the two-fold degeneracy of the spectrum is conserved for all angles $\varphi$, indicating that a $\cos(2\theta)$ potential is equally preserved. 
The instantaneous eigenstates of \cref{eq: circuit hamiltonian} are also compared against those of the ideal case where $\cos(\hat \theta-\hat \phi)$ is replaced by the ideal interaction $\sin(\hat \theta)\sin(\hat \phi)$ for different angles $\varphi$ in \cref{app: numerics}. As observed in \cref{fig: zak patch rotation vs ideal} b) and \cref{fig:Zak patch rotation} b), both models  conserve the desired double-well potential structure responsible for the protection mechanisms of the $\cos(2\hat\theta)$ qubit.

\begin{figure}[t!]
    \centering
    \includegraphics[width=\textwidth]{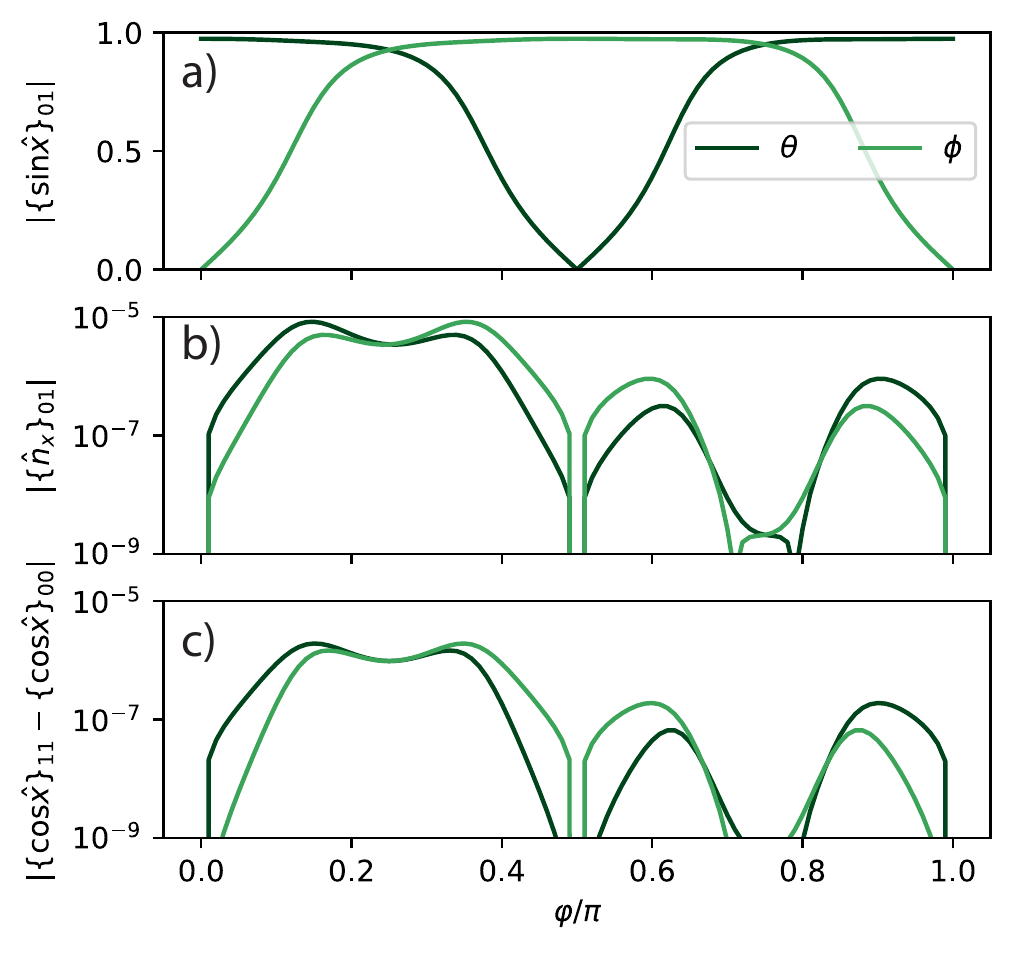}
    \caption{a) $0-1$ matrix elements of the sine operators resulting from flux offsets in the interferometers and resulting in bit-flips. b) $0-1$ matrix elements of the charge operators responsible for bit-flips in presence of low-frequency charge noise. c) Frequency shifts resulting from either frequency flux noise or charge noise in the junction transparency yielding cosine terms.}
    \label{fig:coherence_times}
\end{figure}

A realistic circuit would also be subjected to external charge and flux noise (see \cref{app: circuit implementation}). To leading order, the relevant noise operators are the charge operator as well as the cosine and sine phase operators of each mode. In particular, we observe that both the matrix elements of the charge operators and cosine phase operators remain exponentially suppressed during the gate's adiabatic evolution as expected from preserving a double-well structure in the total potential at all times, see \cref{fig:coherence_times}. Similarly, the sine phase operators remain similar to logical $X$ operators at all times. As a result, dephasing from charge noise, coming from either charge operators or cosine terms due to the voltage tunability of the junctions, is exponentially suppressed. As already noted in Ref.~\cite{Larsen2015}, the matrix element of the  sine terms resulting from low-frequency flux noise are the main source of error for the $\cos(2\theta)$ qubit, see \cref{fig:coherence_times}~c). These correspond to $X$-type errors. Importantly and in analogy to the noise-bias preserving rotation in cat qubits~\cite{Puri2020}, the adiabatic evolution does not yield $Z$-type errors. 

\emph{Numerical results for the $Z$ gate---} In what follows, we account for the time dependence in the angle $\varphi\to\varphi(t)$ with the boundary conditions $\varphi(0)=0$ and $\varphi(T)=\pi$ such as to implement a logical $Z$ gate where the two potential wells in the $\theta$ mode are effectively swapped at time $t=T$. As discussed in \cref{app: time simulations}, constraints on $\varphi(t)$ can be derived from the adiabatic theorem~\cite{Berry_2009} and used to optimize both the gate time and fidelity.

\begin{figure*}[t!]
    \centering
    \includegraphics[width=.8\textwidth]{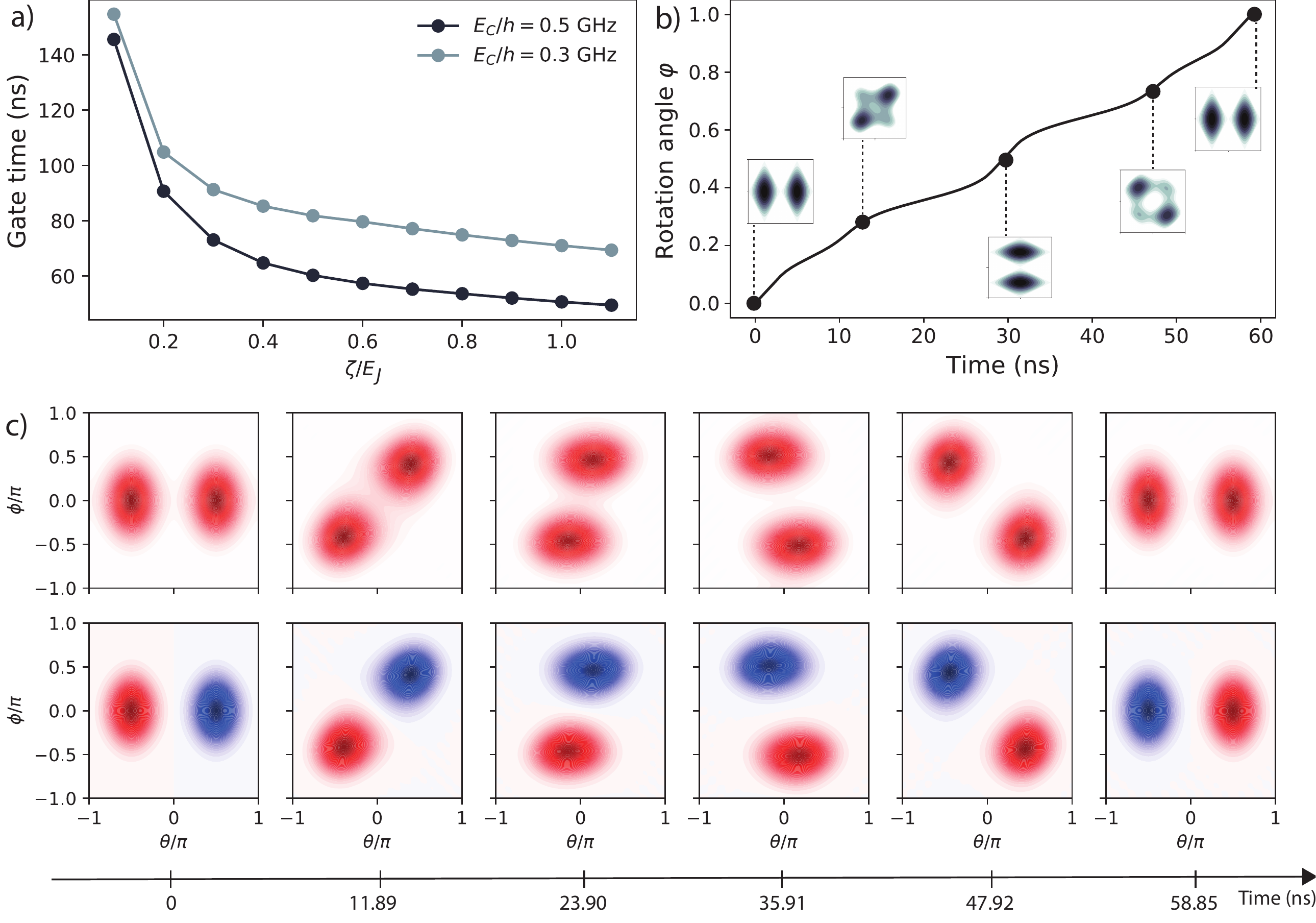}
    \caption{a) Gate time as a function of $\zeta/E_J$ (see \cref{eq: circuit hamiltonian} for definitions) with $E_J/h=\alpha/h=\beta/h=20$ GHz for two different values of the charging energy $E_C$. The fidelity is above $99.95\%$ for all points. b) Example of an optimized time-dependent rotation angle $\varphi(t)$ yielding a $59$ ns gate with $99.97\%$ fidelity for $E_C/h = 0.4$ GHz. The potential energy in 2D phase space for specific times (black dots) are also shown. Details of the pulse optimization can be found in \cref{app: time simulations}. c) Time-evolution of the even (first row) and odd (second row) $Z$ states in the 2D phase basis with the optimized pulse in b). The time-evolution is performed using the charge basis (see \cref{app: time simulations} for details).}
    \label{fig:noise bias gate optimization}
\end{figure*}

In absence of decoherence, we numerically simulate evolution under the Hamiltonian of \cref{eq: circuit hamiltonian} and find that it is possible to realize a $Z$ gate in less than $100$ ns gate with an average gate fidelity~\cite{NIELSEN2002} above 99.96\%, see \cref{fig:noise bias gate optimization} a). The fidelity is computed between the target $Z$ gate unitary and the propagator ${\hat{\mathcal{P}} \cdot \mathcal{T}\exp\left(-i\int_0^t d\tau \hat H[\varphi(\tau)]\right) \cdot \hat{\mathcal{P}}}^\dagger$, where $\hat {\mathcal{P}}$ is the projector onto the logical subspace and $\hat H$ is defined in \cref{eq: circuit hamiltonian}. The shape of $\varphi(t)$ was optimized to reduce leakage during the adiabatic process (see \cref{app: time simulations}). An example of $\varphi(t)$ is shown in \cref{fig:noise bias gate optimization} b) with a time evolution of initial even (first row) and odd (second row) $Z$ states illustrated in c) resulting in a $59$ ns gate with $99.97\%$ fidelity. The methods used for the numerical simulations are detailed in \cref{app: time simulations}. Optimal control techniques can be used to further improve the fidelity.

The limitation on the gate fidelity is better appreciated by considering the non-adiabatic correction~\cite{Berry_2009} (see \cref{app: time simulations}) which is approximately proportional to the time derivative of the effective low-energy model in \cref{eq: low energy H} 
\begin{equation}
\begin{split}
    \frac{d\hat H}{dt} \approx  & -\dot \varphi \beta  \left(2\pi\hat \Phi/d \right)\sin(2\pi\hat\Theta/d) - \dot \varphi\alpha (d \hat \Theta/\pi) \sin(d\hat \Phi/\pi),
\end{split}
\end{equation}
where we assumed the time-derivative of $d$ to be negligible. The fidelity of the gate is maximized when the matrix elements $d\hat H/dt$ involving the instantaneous logical states are small relative to the instantaneous transition frequencies. In the small impedance limit $E_C/\alpha,E_C/\beta \ll 1$, the main source of error takes the form of a beamsplitter interaction between the two modes $\Theta$ and $\Phi$ and can be reduced by increasing the detuning between $\omega_{02}^\Theta(\varphi)$ and $\omega_{01}^\Phi(\varphi)$, where $\omega_{0n}^X(\varphi)$ is the instantaneous $n$th energy level of mode $X$ at angle $\varphi$. Further optimization of the circuit parameters is left for future work.

\emph{Conclusion---} We have introduced an approach to realize a protected gate on a a $\cos(2\theta)$ qubit by coupling it to a second $\cos(2\theta)$ qubit such as to simultaneously preserve the protection mechanisms and realize fast high-fidelity gates.  The ancillary mode allows for an effective beamsplitter operation to take place and a $Z$ gate is realized by adiabatic manipulation of the potential energies of the two qubits, an approach which takes inspiration from gates in the Kerr-cat qubit. The scheme proposed here could in principle be realized with any protected qubit having a double-well potential by supplementing with a second mode. 
A gate protocol such as the one proposed here could render control in protected qubits more accessible. 

\section*{Acknowledgments}

We are grateful to Andras Gyenis, Simon Lieu and Ross Shillito for fruitful discussions and feedback on the manuscript. This research was funded in part by NSERC, the Minist\`ere de l’\'Economie, de l’Innovation et de l'\'Energie du Qu\'ebec, the Canada First Research Excellence Fund, and the U.S. Army Research Office grant No.~W911NF2210042.

\bibliography{refs.bib}

\newpage

\onecolumngrid

\appendix

\section{$\cos(2\theta)$ qubit}
\label{app:cos(2theta) qubit}

The Andreev bound state (ABS) energy of a single superconducting-semiconducting junction (e.g.~a nanowire junction) is~\cite{Weber2018,Larsen2015,Kringhoj2018,Marcus2019,Larsen2020,Vries2021,Lee2015,Haque2021,Schrade2022,Leroux2022}
\begin{equation}
    \varepsilon(V,\hat\theta) = - \Delta \sum_j \sqrt{1 - T_j(V) \sin^2(\hat\theta/2)}, \label{eq:ABS energy}
\end{equation}
where $\Delta$ is the superconducting gap, $T_j(V)$ is the transmission probability of the $j$th channel at the gate voltage $V$, and $\hat\theta$ is the phase across the junction. \Cref{eq:ABS energy} can be expanded in $\hat\theta$ harmonics as 
\begin{equation}
    \varepsilon(V,\hat\theta) = - \sum_{m=1}^\infty  (-1)^{m-1}E_{Jm}(V) \cos\left(m \hat\theta\right), \label{eq:ABS energy Fourier}
\end{equation}
where we have defined the positive-defined energies 
\begin{equation}
    E_{Jm}(V) = \Delta \sum_{n=m}^\infty  2 \begin{pmatrix} 1/2\\ n \end{pmatrix}\begin{pmatrix}2n \\ n-m \end{pmatrix} \sum_i \frac{(-1)^{n+1}T_i^n(V)}{4^n}.
\label{eq:effective EJ}
\end{equation}
To simplify the presentation, the derivation of \cref{eq:ABS energy Fourier} is presented in \cref{app:Expansion of the Andreev bound state energy}.

Using the above result, the Hamiltonian of the circuit of \cref{fig:cos2theta} a) which contains two semiconducting junctions reads
\begin{equation}
\begin{split}
    \hat H_{\cos(2\theta)} & \approx 4 E_C\hat n_\theta^2  - \alpha(V_1,V_2,\phi_\mathrm{ex}) \cos(\hat\theta) + \beta(V_1,V_2,\phi_\mathrm{ex}) \cos(2\hat \theta) + \epsilon(V_1,V_2,\phi_\mathrm{ex}) \sin(\hat\theta), 
\end{split}
\end{equation}
where $\hat \theta$ is the phase operator, $\hat n_\theta$ the number operator and $E_C$ the charging energy. The amplitudes of the potential energy terms take the form
\begin{align}
    \alpha(V_1,V_2,\phi_\mathrm{ex}) &= E_{J1}^1(V_1)+ E_{J1}^2(V_2) \cos(\phi_\mathrm{ex}) \\
    \beta(V_1,V_2,\phi_\mathrm{ex}) &= E_{J2}^1(V_1) + E_{J2}^2(V_2) \cos(2\phi_\mathrm{ex}) \\
    \epsilon(V_1,V_2,\phi_\mathrm{ex}) &= E_{J2}^2(V_2) \sin(\phi_\mathrm{ex}),
\end{align}
where $E_{Jk}^i(V_i)$ is the amplitude of the $k$th harmonic of the Andreev bound state energy of the $i$th junction gated with a voltage $V_i$, and $\phi_\mathrm{ex}$ is the external flux bias.

Away from the flux sweet spot $\phi_\mathrm{ex}=\pi$, the two potential wells become asymmetric due to the $\sin(\theta)$ term. For the remaining of this work we focus on the case $\phi_\mathrm{ex}=\pi$ to simplify the discussion. We note, however, that $\epsilon$ can be tuned to implement $Z$ gates  and to facilitate state preparation in the desired potential well.

\section{Expansion of the ABS energy}
\label{app:Expansion of the Andreev bound state energy}

Taylor expanding \cref{eq:ABS energy} in the transmission probabilities $T_i(V)$ yields
\begin{equation}
    \varepsilon(V,\hat\theta) = - \Delta \sum_i  \sum_{n=0}^\infty \begin{pmatrix} 1/2\\ n \end{pmatrix} (-1)^n T_i^n(V) \sin^{2n}(\hat\theta/2).
\end{equation}
This expression can be put in a more useful form by using the binomial expansion of ${4^n\sin^{2n} \theta = (-1)^n\left(e^{i\hat\theta}-e^{-i\hat\theta}\right)^{2n}}$,
\begin{equation}
\begin{split}
    4^n\sin^{2n}(\hat\theta/2) & =\sum_{m=0}^{2n} \begin{pmatrix}2n \\ m \end{pmatrix}  (-1)^{n+m} e^{i (2n-m)\hat\theta}e^{-im\hat\theta} 
    =\begin{pmatrix}2n \\ n \end{pmatrix} + 2\sum_{m=1}^{n} \begin{pmatrix}2n \\ n-m \end{pmatrix}  (-1)^{m} \cos(m\hat\theta),
\end{split}
\end{equation}
where we used the change of variables $m\to n-m$. Ignoring the constant offset, we find that
\begin{equation}
\begin{split}
    \varepsilon(V,\hat\theta) = - \Delta \sum_i  \sum_{n=0}^\infty \sum_{m=1}^{n} (-1)^{n+m}2\begin{pmatrix} 1/2\\ n \end{pmatrix}  \begin{pmatrix}2n \\ n-m \end{pmatrix} \frac{T_i^n(V)}{4^n} \cos(m\hat\theta),
\end{split}
\end{equation}
which we can reorder as  
\begin{equation}
\begin{split}
    \varepsilon(V,\hat\theta) = - \Delta \sum_i  \sum_{m=1}^{\infty}\sum_{n=m}^\infty (-1)^{n+m}2\begin{pmatrix} 1/2\\ n \end{pmatrix}  \begin{pmatrix}2n \\ n-m \end{pmatrix}\frac{T_i^n(V)}{4^n} \cos(m\hat\theta)
\end{split}
\end{equation}
to recover \cref{eq:ABS energy Fourier}. 

\section{Circuit implementation}
\label{app: circuit implementation}

The total Hamiltonian of the circuit in \cref{fig:Zak patch rotation}~a) has the form $\hat H  = \hat H_\theta + \hat H_\phi + \hat H_{\theta-\phi}$ where 
\begin{align} 
    \hat H_x = 4 E_{Cx} (\hat n_x-n_{\mathrm{ex},x})^2 - \alpha_x(V_{1x},V_{2x},\phi_{\mathrm{ex},x}) \cos(\hat x)+\beta_x (V_{1x},V_{2x},\phi_{\mathrm{ex},x})\cos(2\hat x)+\epsilon_x(V_{1x},V_{2x},\phi_{\mathrm{ex},x}) \sin(\hat x)
\end{align}
is the $\cos(2\theta)$ qubit Hamiltonian for mode $x \in \{\theta,\phi\}$ and where we have used the results of \cref{app:cos(2theta) qubit,app:Expansion of the Andreev bound state energy}. Moreover, the last term of $\hat H$ takes the form
\begin{align}
    \hat H_{\theta - \phi} = g \hat n_\theta \hat n_\phi  -\alpha_g(V_{1g},V_{2g},\phi_{\mathrm{ex},g}) \cos(\hat \theta-\hat\phi)+\beta_g (V_{1g},V_{2g},\phi_{\mathrm{ex},g})\cos(2\hat \theta-2\hat \phi)+\epsilon_g(V_{1g},V_{2g},\phi_{\mathrm{ex},g}) \sin(\hat \theta - \hat \phi),
\end{align}
This represents the Hamiltonian of the coupler which is also a semiconductor interferometer mediating a capacitive coupling $g$ between the two $\cos(2\theta)$ qubits as well as an inductive interaction through the semiconductor junctions. Here we consider $\phi_{\mathrm{ex},x}=\pi$ for $x \in \{\theta,\phi,g\}$, the small transmission limit where $\beta_g \to 0$, and the weak coupling limit $g\to 0$. Controlling $V_{1x}$ and $V_{2x}$ allows us to implement the Hamiltonian in \cref{eq: circuit hamiltonian}. 

\section{Low energy model}
\label{app: low energy}

The global two-fold degenerate minima of the potential energy of \cref{eq: circuit hamiltonian} are approximately defined on opposite points on the edges of a square of width $\pi$
\begin{equation}
    (\theta,\phi) = \pm 
    \begin{cases}
        \left(\pi/2,\pi\tan(\varphi)/2\right), & 0 \leq \varphi \leq \pi/4, \\
        \left(\pi\cot(\varphi)/2,\pi/2\right), & \pi/4 < \varphi < 3\pi/4,\\
        \left(-\pi/2,-\pi\tan(\varphi)/2\right), & 3\pi/4 \leq \varphi \leq \pi. 
    \end{cases} \label{eq: minima}
\end{equation}
The instantaneous distance between the global minima is therefore 
\begin{equation}
    d(\varphi) = \pi |\sec(\min (\varphi,|\pi/2-\varphi|,|\pi-\varphi|)|.
\end{equation}

The square can be parameterized into a circle with varying diameter $d(\varphi)$, the distance between the two global minima. We can therefore write an effective low-energy Hamiltonian
\begin{equation}
    \hat H \approx 4 E_{C} \hat n_\Theta^2 + \beta \cos[2\pi\hat \Theta/d(\varphi)] \\
    + 4 E_{C}\hat n_\Phi^2 - \alpha \cos[d(\varphi)\hat \Phi/\pi], \label{eq: effective Hamiltonian low energy}
\end{equation}
where $\hat \Theta = \cos(\varphi)\hat \theta + \sin(\varphi)\hat \phi$ is defined along the radius of the rotation and $\hat \Phi = \cos(\varphi)\hat \phi - \sin(\varphi)\hat \theta$ is its perpendicular coordinate. To preserve the $2\pi$ periodicity of the transformed coordinates we can also renormalize $\hat \Theta \to d(\varphi) \hat \Theta/\pi$ and $\hat \Phi \to \hat d(\varphi)\Phi/\pi$. \Cref{eq: effective Hamiltonian low energy} gives a qualitatively good description of the adiabatic path in that it accurately follows the global minima.

\Cref{eq: effective Hamiltonian low energy} does not, however, captures all the low-order effects of the term $-(\zeta/2)\sin(2\varphi)\cos(\hat\theta-\hat \phi)$ of \cref{eq: circuit hamiltonian}. Beyond reducing the four-fold degeneracy to a two-fold degeneracy, this term also leads to squeezing along either $\hat \Theta$ or $\hat \Phi$ during the rotation, as observed in \cref{fig:Zak patch rotation}~b). This fact is better appreciated by writing $-(\zeta/2)\sin(2\varphi)\cos(\hat\theta-\hat \phi)$ in terms of the rotating coordinates 
\begin{align}
\begin{split}
    -(\zeta/2)\sin(2\varphi)\cos(\hat\theta-\hat \phi)& 
= -(\zeta/2)\sin(2\varphi)\cos\left(\left[\cos(\varphi)-\sin(\varphi)\right]\hat \Theta - \left[\cos(\varphi)+\sin(\varphi)\right]\hat \Phi\right).
\end{split}\label{eq: cos term}
\end{align}
Taylor expanding \cref{eq: cos term} to second order near $(\hat \Theta,\hat\Phi) = (0,0)$ reveals that \cref{eq: cos term} can be approximated with 
\begin{align}
\begin{split}
    -(\zeta/2)\sin(2\varphi)\cos(\hat\theta-\hat \phi)&\approx (\zeta/4)\sin(2\varphi)\left[\cos(\varphi)-\sin(\varphi)\right]^2 \hat \Theta^2+(\zeta/4)\sin(2\varphi)\left[\cos(\varphi)+\sin(\varphi)\right]^2\hat \Phi^2.
\end{split}\label{eq: cos term approx}
\end{align}
We can therefore correct \cref{eq: effective Hamiltonian low energy} with 
\begin{equation}
    \hat H \approx 4 E_{C} \hat n_\Theta^2 + (1-r(\varphi)) \beta \cos[2\pi\hat \Theta/d(\varphi)] \\
    + 4 E_{C}\hat n_\Phi^2 - (1+s(\varphi))\alpha \cos[d(\varphi)\hat \Phi/\pi], \label{eq: effective Hamiltonian low energy corrected}
\end{equation}
where we defined 
\begin{align}
    & r(\varphi) = \frac{\zeta \sin(2\varphi)}{2\beta} \left[\frac{d(\varphi)}{2\pi}\right]^2\left[\cos(\varphi)-\sin(\varphi)\right]^2, \\
    & s(\varphi) = \frac{\zeta \sin(2\varphi)}{2\alpha} \left[\frac{\pi}{d(\varphi)}\right]^2\left[\cos(\varphi)+\sin(\varphi)\right]^2.
\end{align}

This distortion observed in \cref{fig:Zak patch rotation} b) is however not captured by either \cref{eq: effective Hamiltonian low energy} or \cref{eq: effective Hamiltonian low energy corrected} and results from interactions between $\Theta$ and $\Phi$. Even though they don't significantly impact the global minima, they renormalize the matrix elements of the charge operators and cosine phase operators, as shown in \cref{fig:coherence_times} b-c). In particularly, they become non-zero during the rotation but remain exponentially suppressed due to the disjoint support of the eigenstates. 

\begin{figure}[t!]
    \centering
    \includegraphics[width=\textwidth]{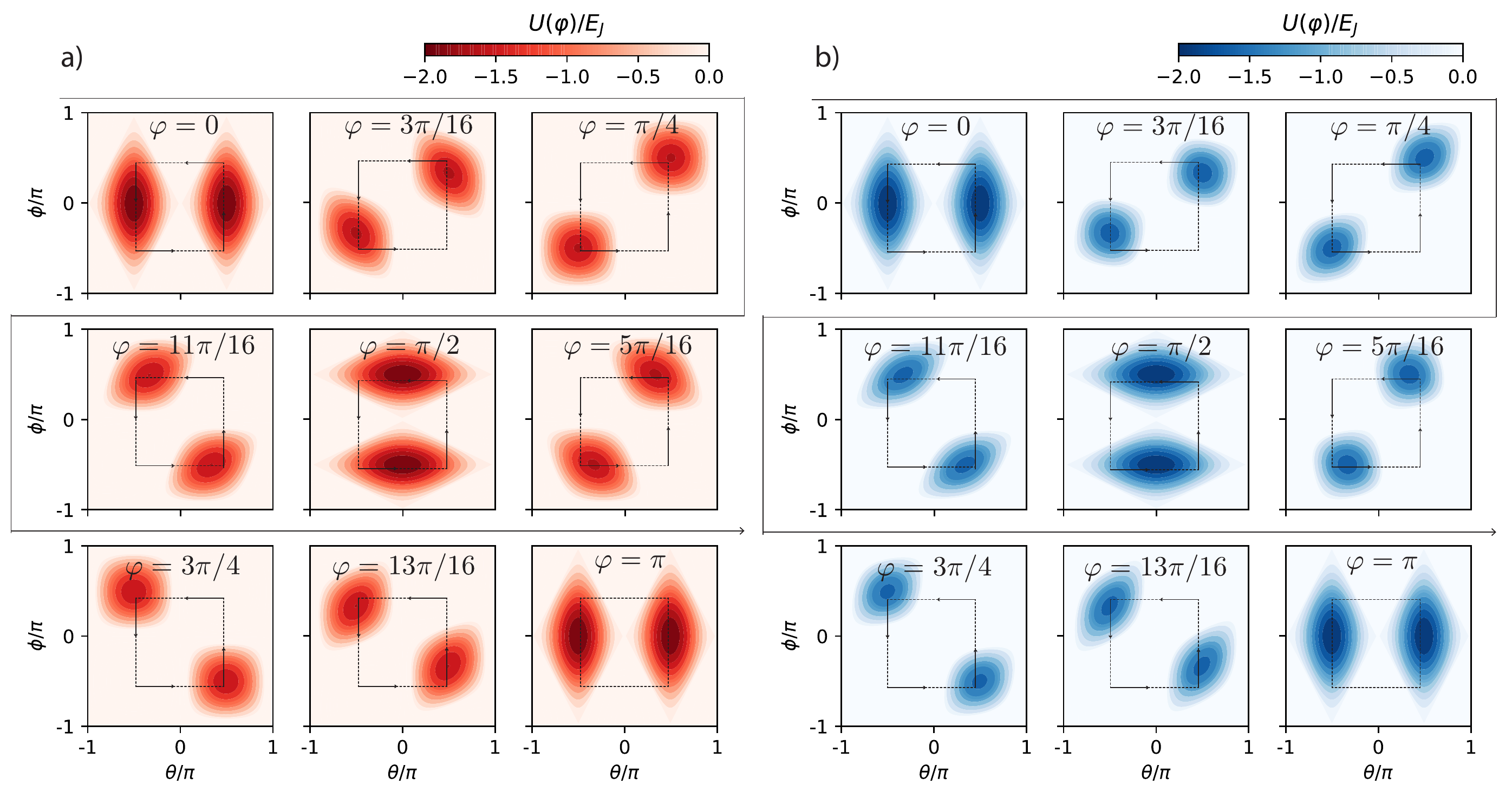}
    \caption{a) Reproduction of \cref{fig: zak patch rotation vs ideal} b) with the potential energy of the low-energy model in \cref{eq: effective Hamiltonian low energy}. This low-energy model ignores the effect of the  $\cos(\hat\theta)\cos(\hat\phi)$ term and ignores any interaction between rotating coordinates $\Theta$ and $\Phi$ which result in some additional distortion and diffusion in \cref{fig: zak patch rotation vs ideal} b). b) Reproduction of \cref{fig:Zak patch rotation} b) with the potential energy of \cref{eq: effective Hamiltonian low energy corrected}. This model now includes the leading order effects of the $\cos(\hat\theta)\cos(\hat\phi)$ term resulting in squeezing along either $\hat \Theta$ or $\hat \Phi$ and in asymmetry of the potential energy during the rotation. Any interaction between $\hat\Theta$ and $\hat \Phi$ is still ignored within the low-energy approximation.}
    \label{fig:effective_low_energy_models}
\end{figure}

\section{Comparison of the $\sin(\theta)\sin(\phi)$ and $\cos(\theta-\phi)$ models}
\label{app: numerics}

\begin{figure}[t!]
    \centering
    \includegraphics[width=\textwidth]{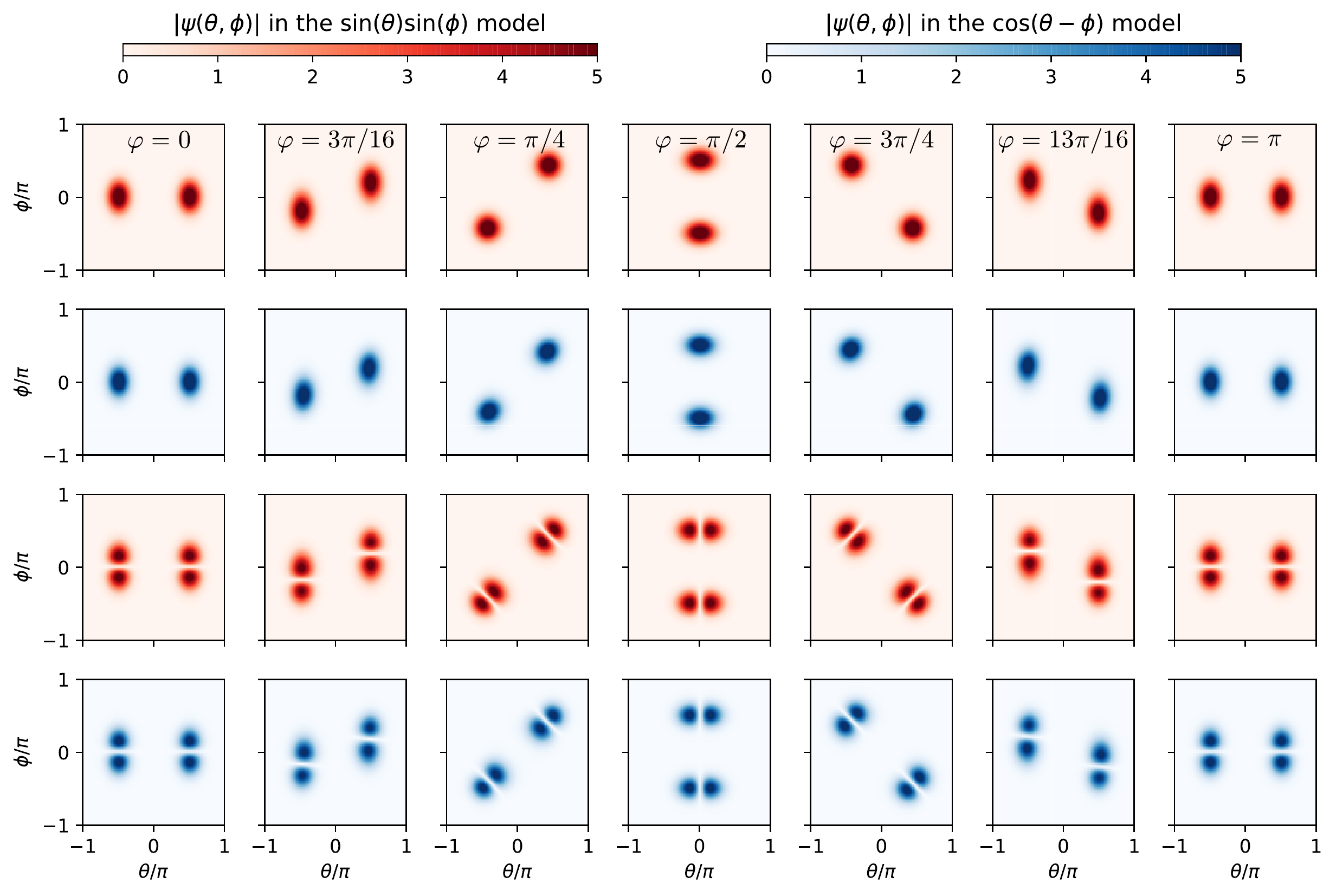}
    \caption{Instantaneous ground state and second excited state as a function of the rotation angle $\varphi$, in phase basis, with the proposed circuit implementation in \cref{eq: circuit hamiltonian} (in red) and the ideal version where ${\cos(\hat \theta-\hat \phi) \to \sin(\hat \theta)\sin(\hat \phi)}$ (in blue). Here the diagonalization is realized in charge representation for both modes. The eigenstates are then Fourier transformed to return the phase representation. Here ${E_C/h = 100}$ MHz and ${\alpha/h=\beta/h=\zeta/h = 20.0}$ GHz.}
    \label{fig:eigenstates}
\end{figure}

\begin{figure}[t!]
    \centering
    \includegraphics[width=\textwidth]{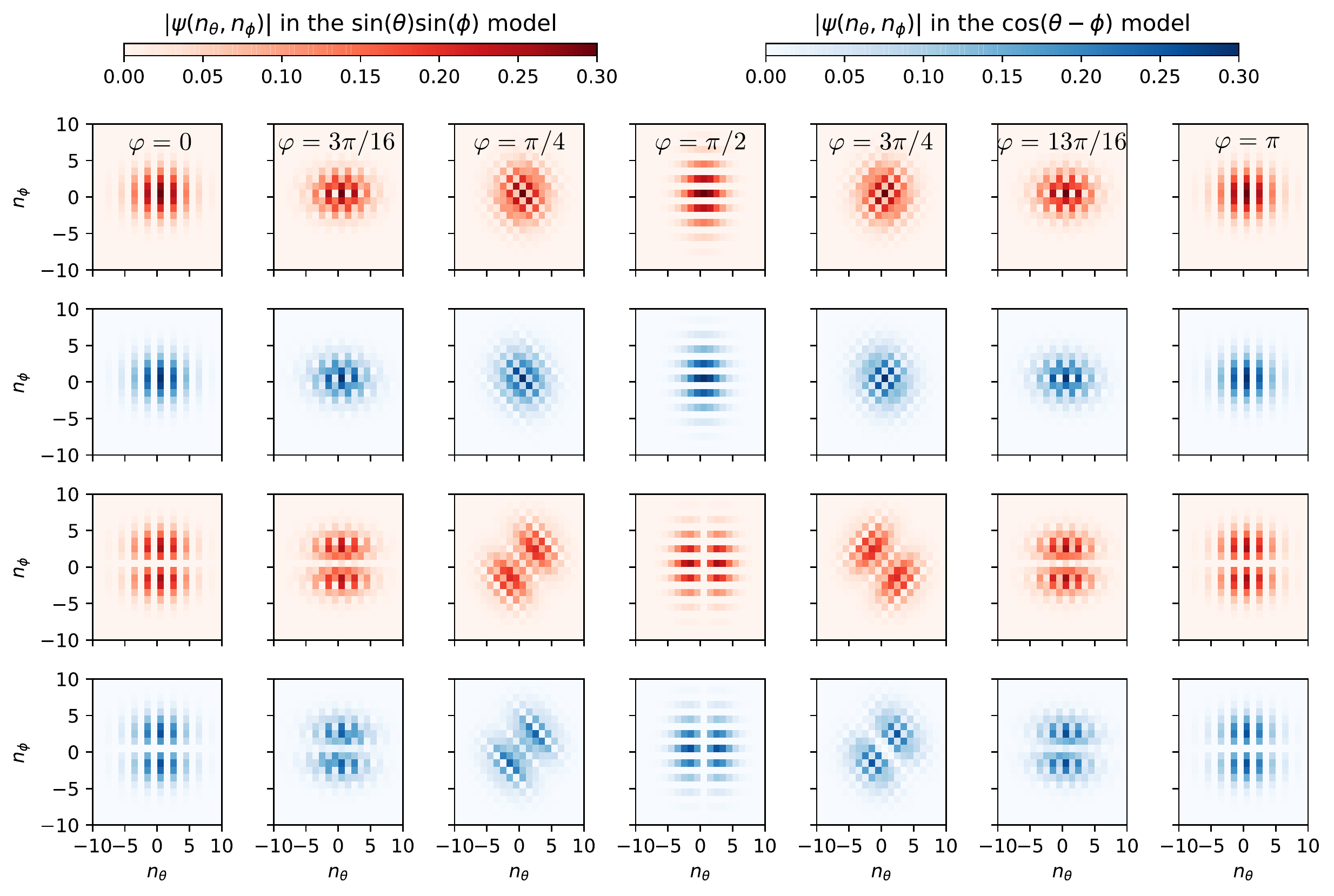}
    \caption{Same as \cref{fig:eigenstates} but in the charge basis.}
    \label{fig:eigenstates_charge}
\end{figure}

The ground and second excited states of both the $\sin(\theta)\sin(\phi)$ and $\cos(\theta-\phi)$ models are shown in 2D phase space in \cref{fig:eigenstates}. These results are obtained from numerical diagonalization of the Hamiltonians in the charge basis. The conversion to phase basis is done via a Fourier transform in the two coordinates. The  eigenstates are also shown in the charge basis in \cref{fig:eigenstates_charge}. Since the phase coordinates of the two $\cos(2\theta)$ qubits are compact the states have discrete charge states.

We remark that the low-energy eigenstates in both models share similar features. The key difference arises when comparing $\varphi=\pi/4$ and $\varphi = 3\pi/4$, especially for the second excited state. There, the states are rotated by $\pi/2$ locally near each global minimum in the 2D phase space (see \cref{fig:eigenstates}) for $\varphi=3\pi/4$ but not $\varphi=\pi/4$. This asymmetry results from the $\cos(\theta)\cos(\phi)$ term present in the $\cos(\theta-\phi)$ interaction. Even though this term slightly renormalizes the energies and matrix elements of the logical states, it does not prevent the formation of a double well potential at all angles $\varphi$ as shown in \cref{fig:eigenstates}. In other words, its impact on the protection of the qubit is negligible.

\section{Time-dependent simulations}
\label{app: time simulations}

We promote the angle $\varphi\to\varphi(t)$ to an instantaneous angle of rotation at time $t$ with the boundary conditions $\varphi(0)=0$ and $\varphi(T)=\pi$ for a total gate time $t=T$. 

In the spirit of Ref.~\cite{Berry_2009}, we introduce the time-dependent unitary ${\hat{\mathcal{A}}[\varphi(t)] =  \sum_n \ketbra{\psi_n[\varphi(t)]}{\psi_n(0)}}$, 
where $\ket{\psi_n[\varphi(t)]}$ is the instantaneous eigenstates of $\hat H[\varphi(t)]$ with energies $\omega_n[\varphi(t)]$. The Hamiltonian can then be expressed as 
\begin{equation}
\begin{split}
    &\hat H_\mathcal{A}[\varphi(t)] = \sum_{n=0}^\infty \omega_n[\varphi(t)] \ketbra{\psi_n(0)}{\psi_n(0)} - i \sum_{n,m=0}^\infty \ket{\psi_n(0)} \matrixel{\psi_n[\varphi(t)]}{\frac{d}{dt}}{\psi_m[\varphi(t)]}\bra{\psi_m(0)},
\end{split}
\end{equation}
where
\begin{equation}
    \matrixel{\psi_n[\varphi(t)]}{\frac{d}{dt}}{\psi_m[\varphi(t)]} = \dot{\varphi}\frac{\matrixel{\psi_n[\varphi(t)]}{\frac{d\hat H}{d\varphi}}{\psi_m[\varphi(t)]}}{\omega_n[\varphi(t)]-\omega_m[\varphi(t)]} \label{eq: <psi_n|dt|psi_m>}
\end{equation}
for ${\omega_n[\varphi(t)]\neq \omega_m[\varphi(t)]}$. 

For the adiabatic theorem to hold, the magnitude of these matrix elements must be much smaller than ${|\omega_n[\varphi(t)]-\omega_m[\varphi(t)]|}$. In this case, the system remains in the $n$th instantaneous eigenstate. Quantum transitionless driving~\cite{Berry_2009} can in principle be used 
to reduce the amplitude of these transitions to speed up the gate. 

We optimize the shape of $\varphi(t)$ by bounding ${10^ {-3}|\omega_n[\varphi(t)]-\omega_m[\varphi(t)]|}$ for $n=0,1$ at all times.

The time-dependent simulations are done in the charge basis for both compact modes $\theta$ and $\phi$ with sparse matrices by solving the time-dependent Schr\"odinger equation with the Hamiltonian in \cref{eq: circuit hamiltonian}. The states are then converted to phase space using Fourier transforms.

\end{document}